\documentclass[preprint,eqsecnum,aps]{revtex4}
\usepackage{graphicx}
\usepackage{epsfig}
\usepackage{amsfonts}
\usepackage{amssymb}
\begin{document}
\title{Semiclassical description of a sixth order quadrupole boson Hamiltonian}

\author{F. D. Aaron $^{a)}$ and A. A. Raduta $^{a),b)}$}

\address{$^{a)}$ Department of Theoretical Physics and Mathematics,
Faculty of Physics, Bucharest University, POBox MG11, Romania }

\address{$^{b)}$ Department of Theoretical Physics, Institute of Physics and
Nuclear Engineering, Bucharest, POBox MG6, Romania}

\begin{abstract}
A sixth order quadrupole boson Hamiltonian is treated through a time dependent variational
principle approach choosing as trial function a coherent state with respect to
zeroth $b^{\dagger}_0$ and second $b^{\dagger}_2+b^{\dagger}_{-2}$ components
of the quadrupole bosons. The coefficients involved in the model Hamiltonian
are chosen so that the classical effective potential energy term has two distinct minima.
The classical system is described by two degrees
of freedom and  has two constants of motion. The equation of motion for the
radial coordinate is analytically solved and the resulting trajectories are 
extensively  studied.
One distinguishes three energy regions exhibiting different types of trajectories.
When one passes from  the region characterized by two wells to the region of
energies higher than the maximum value of the effective potential the
trajectories period exhibits a singularity which reflects in fact a
phase transition. The classical trajectories are quantized by a constraint
of the integral action similar to the well known Bohr-Sommerfeld quantization
condition.
The semiclassical spectra corresponding to the two potential wells have specific
properties.
The tunneling process through the potential barrier is also studied.
It is a remarkable fact that the transmission coefficients exhibit jumps in
magnitude when the angular momentum acquires certain values. One suggests that
for such angular momenta the rates of alpha or heavy cluster emissions might be
very large.
\end{abstract}

\maketitle

\section{Introduction}
Description of spectroscopic properties for a  many body nuclear system
is one of the most interesting aims of theoretical nuclear studies. Some of
such properties are obtainable by using only few degrees of freedom.
Thus the liquid drop model \cite{Bo}
tries to describe the basic quadrupole collective properties by using a harmonic
liquid drop whose surface vibrates with small amplitudes around an equilibrium
spherical shape.
Few years later another two solvable simple formalisms have been launched namely
the Jean-Wilets model for gamma unstable nuclei \cite{JeWi} and the Davydov
Filippov (DV) model of
triaxial rigid rotor \cite{DaFi}. Although these models are not able to describe
the complex details of nuclear spectra they have the big merit to define
the framework and to place reference landmarks in the field.
Indeed,  many of the improvements proposed interpret their results by considering the
departure from the reference picture provided by the simple models mentioned
before.
Thus, the first phenomenological model which goes beyond both the liquid drop
and DF models is the vibration rotation model \cite{FaGr} where the drop surface may oscillate
around a static deformed shape. The anharmonic terms in the quadrupole shape coordinates are
added to the liquid drop for the first time by Greiner and Gneuss \cite{GnGr}.
This anharmonic model was extensively used to describe various types of spectra
, gamma unstable, gamma stable, triaxial shapes in several publications \cite{Hess}.
The drawback of this model is the large number of parameters involved although they depend
smoothly on the atomic mass number.
A much smaller number of parameters is used by the coherent state
model (CSM)\cite{Rad1}
which treats three interacting rotational bands within a restricted collective space
defined through angular momentum projection technique, from three deformed states: one of
them is describing the deformed ground
state and is of an axial symmetric  coherent function and the other two are simple polynomial excitations
of the ground state determined by requiring the three states to be mutually
orthogonal.
CSM exploits the semi-classical features of coherent states which results in obtaining a suitable
description for states of high angular momentum. It also provides a unified
description of near vibrational, transitional
and of well deformed nuclei.
Another interesting phenomenological model is the interacting boson approximation
which uses besides the quadrupole bosons a monopole one \cite{Ia}. The model Hamiltonian is
conserving the total number of bosons which as a matter of fact is limited to
half the number of
valence nucleons from the nucleus under consideration. By construction, the
Hamiltonian accounts for one of the symmetries O(6), SU(5) and SU(3).
Consequently, the nuclear states are just the irreducible representations for
these extreme symmetries. However, to describe nuclei departing from
these symmetries additional terms are to be added which blurs the primary
beauty of the model.

The phenomenological descriptions involving anharmonic boson Hamiltonians make use
either of approximate methods or of diagonalization procedures to get the
desired eigenvalues.
In the latter case the procedure is difficult to be applied when we are close
to the picture where the results are very slowly convergent against increasing
 the basis dimension.
On the other hand, when an approximate approach is used one does not know
how reliable it is . In other words, we don't know how
far the obtained result is from the exact one.

The same  difficulties are encountered also in microscopic formalisms.
The microscopic theories interpret the experimental data in terms of
single particle motion. Thus, the collective motion is
defined in terms of individual degrees of freedom. The random phase approximation (RPA)
or boson expansions are defining boson operators in terms of bifermionic operators
and then the Hamiltonian and transition operators are written as series of the
RPA bosons \cite{Ze,Mar,Kle}. The convergence properties of the boson series and the magnitude of the
ignored "rest", when a truncation is performed, can be evaluated only for
exactly solvable models. The most known solvable microscopic models are those
of Moszkowski \cite{Moz}, Lipkin-Meshkov \cite{LiMe}
and one level pairing Hamiltonian \cite{onelev}.
Applying for example, one of the above mentioned approaches to a solvable model
and comparing
the results with the corresponding exact output one may conclude upon the validity
of the procedure adopted. It should be stressed the fact that using a realistic
Hamiltonian instead of a solvable one, one expects that many of the features
are at least qualitatively preserved.

It is worth mentioning that for phenomenological boson models including high
anharmonicities exact results are lacking.
In a previous publication we studied three  fourth order boson Hamiltonians which are
exactly solvable\cite{Rad2}. For two of them we analyzed the possibility of classifying
the states describing the intrinsic degrees of freedom in rotational bands.

In the present paper we continue the project started in our previous paper and
study semi-classically a sixth order boson Hamiltonian which is exactly solvable.
Results for the exact trajectories, their periods and quantization are presented in an
analytical form. Interesting results for the transmission probability through
a barrier are obtained within the WKB approach.

The formalism and final results are presented according to the following plan.
In Section II, a sextic boson Hamiltonian is chosen.
This is treated within a time dependent variational principle approach in Section III.
The quantization of one periodic coordinate and the semi-classical spectrum is given for
several distinct energy regions in Section IV. The main results and final conclusions are
collected in Section V.
\section{The model sextic boson Hamiltonian}
In the present paper we shall treat semi-classically a sixth order quadrupole
boson Hamiltonian
\begin{equation}
H=A_1\sum_{\mu}b^{\dagger}_{\mu}b_{\mu}+A_2\sum_{\mu}\left
(b^{\dagger}_{\mu}b^{\dagger}_{-\mu}+b_{\mu}b_{\mu}\right)(-)^{\mu}+
A_4{\hat P}^2+A_6\hat {P}^3,
\label{Has}
\end{equation}
where $b^{\dagger}_{\mu}$ ($b_{\mu}$) with $-2\leq \mu \leq 2$,
denotes the creation (annihilation) quadrupole boson operator and
\begin{equation}
\hat {P}=\frac{1}{2}\sum_{\mu}\left( b^{\dagger}_{\mu}+(-)^{\mu}b_{-\mu}\right)
\left(b^{\dagger}_{-\mu}+(-)^{\mu}b_{\mu}\right)(-)^{\mu}.
\label{Pe}
\end{equation}
Writing the high order boson terms, involved in $H$, in a normal order it results that the
model Hamiltonian does not commute with the boson number operator
\begin{equation}
\hat {N}=\sum_{\mu}b^{\dagger}_{\mu}b_{\mu}.
\label{En}
\end{equation}
In what follows we are interested in solving the time dependent variational
principle (TDVP) equations
\begin{equation}
\delta \int_{0}^{t}\langle \Psi|\left(H-i\hbar\frac{\partial}{\partial t'}\right)
|\Psi\rangle dt'=0.
\label{delta}
\end{equation}
When the variational state $|\Psi\rangle$ spans the whole space of the boson states, solving the equation
(2.4) is equivalent to solving the time dependent equation associated to the
model Hamiltonian $H$.
The classical features encountered by $H$ can be described by restricting the space of
$|\Psi\rangle$ to the coherent states:
\begin{equation}
|\Psi\rangle = \rm{exp}\left[z_0b^{\dagger}_0-z^*_0b_0+z_2(b^{\dagger}_{2}+
b^{\dagger}_{-2})-
z^*_2(b_{2}+b_{-2})\right]|0\rangle.
\label{Psi}
\end{equation}
Here the boson vacuum state is denoted by $|0\rangle$. The function
$|\Psi\rangle$  depends on the complex parameters $z_0, z_2$ and their
complex conjugates $z^*_0, z^*_2$.
These parameters play the role of classical phase space coordinates whose equations of motion
are provided by the TDVP equations.

It is convenient to have the classical equations of motion in a canonical form.
In order to touch this goal we perform the following change
of variables
\begin{equation}
q_i=2^{(k+2)/4}{\bf Re}(z_k),\;p_i=2^{(k+2)/4}{\bf Im}(z_k), \; k=0,2,\;\;i=\frac{k+2}{2}.
\label{qi}
\end{equation}
Up to an additive constant, $\frac{35}{4}A_4+\frac{315}{8}A_6$, the classical energy
function, defined as the average value of the model Hamiltonian on the
coherent state $|\Psi\rangle$, has the expression:
\begin{equation}
{\cal H}\equiv \langle\Psi|H|\Psi\rangle=\frac{A'}{2}(p_1^2+p_2^2)+\frac{A}{2}(q_1^2+q_2^2)
+\frac{D}{4}(q_1^2+q_2^2)^2+\frac{F}{6}(q_1^2+q_2^2)^3\;,
\label{Hasro}
\end{equation}
where the following notations have been used:
\begin{eqnarray}
A'&=&A_1-2A_2,\;\; A=A_1+A_2+14A_4+\frac{189}{2}A_6\;,\nonumber\\
D&=&4A_4+54A_6,\;\;F=6A_6.
\label{ADF}
\end{eqnarray}
The nice feature of the classical energy function ${\cal H}$ consists of the fact that it does not
contain powers of momenta higher than two, although we started with a high order
boson Hamiltonian.

It is worth mentioning that similar time dependent approaches are also used for treating
many body systems, when for example a Hartree Fock or  an
RPA approximation \cite{Ring,Rad3} is performed.
 In contradistinction to those formalisms where the classical coordinates, parameterizing the trial wave
function, are small expansion parameters, here no approximation is made in deriving the classical equations of motion
from the variational principle of Eq.(2.4)

The objective of the next section is to study the classical motion of the system
associated to the classical energy function ${\cal H}.$

\section{The classical description}

\bigskip
The classical energy expressed in terms of the polar coordinates $(r,\theta )$
associated to the plane ($q_1,q_2$), looks like:
\begin{equation}
{\cal H}=\frac{\hbar^2}{2A'}(\dot{r}^2+r^2\dot{\theta}^2)+V(r),
\label{hasrond}
\end{equation}
where $V(r)$ is the potential energy which contains in addition to the potential
used in the previous publication, a sextic term:
\begin{equation}
V(r)=\frac{1}{2}Ar^{2}+\frac{1}{4}Dr^{4}+\frac{1}{6}Fr^{6}\;.
\label{Vofr}
\end{equation}
For $r>0$, depending on the coefficients involved, the potential energy
function exhibits either two or no extreme points.
In the first situation the ordering of the maximum and minimum points is decided
by the relative signs of the defining  coefficients A, D and F. Here we study
the case
  $A>0,$ $D<0,$ $F>0$ which defines a potential having first a maximum and then a minimum.
The other ordering situation will be considered in a subsequent work.
Throughout this paper the applications are made with the following values:

\begin{equation}
A=3\; \rm{MeV},\;\;D=-0.4\; \rm{MeV},\;\;F=0.01\;\rm{MeV},\;\;
A^{\prime }=0.0025\; \rm{MeV}.
\label{coeff}
\end{equation}

There are  two constants of the motion,  the energy:
\begin{equation}
{\cal H}=E,
\label{HeqE}
\end{equation}
and  the third component of the angular momentum,
\begin{equation}
{\cal L}_3\equiv \frac{\hbar}{2}(q_1p_2-q_2p_1)= \frac{\hbar ^{2}}{A^{\prime }}r^{2}\dot{\theta}\;.
\label{L3}
\end{equation}
For the sake of completeness we give also the expressions of the other two components of angular momentum
\begin{equation}
{\cal L}_1=\frac{\hbar}{4}(q_1^2+p_1^2-q_2^2-p_2^2),\;{\cal L}_2=\frac{\hbar}{2}
(q_1q_2+p_1p_2).
\label{L1andL2}
\end{equation}
These components generate a classical $SU_c(2)$ algebra with the multiplication operation:
\begin{equation}
\{{\cal L}_i,{\cal L}_k\}=\hbar \epsilon_{ikj}{\cal L}_j\;,
\label{Pobra}
\end{equation}
where $\{,\}$ denotes the Poisson brackets and $\epsilon_{ikj}$ the antisymmetric unit tensor.
${\cal L}_i$ might be obtained as average values on $|\Psi\rangle$ (see Eq. (2.5)) of angular momentum components operators ${\hat L}_i$ acting on a boson 
space \cite{Rad2}. These operators generate a boson $SU_b(2)$ algebra.
The correspondence $[,]\to (1/i)\{,\}$ realizes a homeomorphism of the two
algebras mentioned above.
The existence of the constant of motion ${\cal H}$ is a consequence of the fact the classical
equations of motion are obtained through the variational principle equation (2.4).
On the other hand the equation ${\cal L}_3=const.$ holds since in the quantal picture
the conservation law equation $[H,\hat {L}_3]=0$ takes place.
The classical system has two constants of motion ${\cal H}$ and ${\cal L}_3$ and two degrees of freedom
$r$ and $\theta$. Therefore the system is fully integrable and consequently analytical solutions
for trajectories are expected. As a matter of fact this is one of the aims of the present investigation.

Note that $\frac{\hbar ^{2}}{A^{\prime }}$ plays the role of the  mass
,  $m_{0},$ of a classical non-relativistic particle moving in a central force field.
Also we note that when the sixth power term in $r$ is missing one obtains the classical energy function
used in our previous publication \cite{Rad2}. Moreover the classical energy may
 be viewed as a
counterpart of
a microscopic Hamiltonian including a two body monopole-monopole
interaction. In that case the coordinate $r$ signifies the classical image of a collective
microscopic coordinate \cite{Hugo}.

Eliminating the angular variable in Eq.(\ref{hasrond}) one obtains:
\begin{equation}
\frac{\hbar ^{2}}{2A^{\prime }}\dot{r}^{2}+V_{eff}(r)=E,
\label{rpunct}
\end{equation}
where $V_{eff}$ $(r)$ is the effective potential energy  given by:
\begin{equation}
V_{eff}(r)=\frac{A^{\prime }{\cal L}_{3}^{2}}{2\hbar ^{2}r^{2}}+V(r).
\label{Veff}
\end{equation}
Since ${\cal L}_3$ is the third component of the intrinsic angular momentum,
this is the classical
counterpart of the $K$ quantum number of the liquid drop model. Having this feature in mind
we quantize it by the restriction
\begin{equation}
 {\cal L}_{3}=2L\hbar,\;\;L=0,1,2,...
\label{L3qua}
\end{equation}
Replacing ${\cal L}_3$ by its quantized expression, the effective potential becomes:
\begin{equation}
V_{eff}(L;r)=\frac{2A^{\prime }L^{2}}{r^{2}}+V(r).
\label{VeffL}
\end{equation}
For $L<59$, $V_{eff}(L;r)$ exhibits a two minima shape.
For the critical value $L=59$, the minimum of the left well gets
unified with the maximum of $V_{eff}$ which results in arising an inflection
point. For  $L=25$ and the set of parameters specified by Eq.(3.3), $V_{eff}$ $(L;r)$  was plotted as function of $r$ in Fig. 1.

From Eqs. (3.8), (3.5) and (3.10), one obtains the equations of motion for 
$x=r^2$ and $\theta$, respectively:

\begin{eqnarray}
dt&=&\frac{\hbar }{2}\sqrt{\frac{3}{A^{\prime }F}}\frac{dx}{\sqrt{%
P_{4}(x;E,L)}}, \nonumber\\
d\theta &=&L\sqrt{\frac{3}{A^{\prime }F}}\frac{dx}{x\sqrt{P_{4}(x;E,L)}},
\label{dtdteta}
\end{eqnarray}
where

\begin{equation}
P_{4}(x;E,L)=-x^{4}-\alpha x^{3}-\beta x^{2}-\gamma \left( E\right)
x-\delta (L)
\label{P4}
\end{equation}

with

\begin{eqnarray}
\alpha &=&\frac{3D}{2F}<0,\;\;\beta =\frac{3A}{F}>0,\\ \nonumber
\gamma (E)&=&-\frac{6E}{F}<0,\;\;\delta (L)=12\frac{A^{\prime }L^{2}}{F}>0.
\label{alfa}
\end{eqnarray}

For any $L$ $<59$ there are three energy domains defined as follows (see figure 1 for the case $L=25$):
\begin{eqnarray}
&a)& \rm{UNDER, for}\; E\in \lbrack V_{eff\text{ }\min 2}(L),V_{eff\text{ }%
\min 1}(L)], \nonumber \\
&b)& \rm{BETWEEN, for}\; E\in \lbrack V_{eff\text{ }\min 1}(L),V_{eff\text{
}\max }(L)],       \\  \nonumber
&c)& \rm{OVER, for}\; E\in \lbrack V_{eff\text{ }\max }(L),+\infty ).
\end{eqnarray}
The labels for the three intervals will be hereafter abbreviated by U, B and O respectively
whenever one wants to mention the fact that a given observable characterizes a certain energy region.
The system motion is allowed only for the values of "x" where the polynomial
P acquires positive values. Such allowed intervals are depicted separately for each of the energy regions
U,B and O.
\subsection{Results for an  "U" energy}

In the energy interval U, the motion is possible only in the right well
(see Fig. 1). Indeed, inside U, the equation
\begin{equation}
P_{4}(x;E,L)=0,
\label{P4eq0}
\end{equation}
has two real
\begin{equation}
x_{1}=a(E,L),\;\;x_{2}=b(E,L),\;\;x_{1}>x_{2},
\label{x1x2}
\end{equation}
and two complex conjugate solutions:
\begin{equation}
x_{3}=u(E,L)+iv(E,L)=x_4^*,
\label{x3x4}
\end{equation}
It is conspicuous now that $P_4(x;E,L)\geq 0$ for $x\in [x_2,x_1]$.

For what follows it is useful to introduce the equation
\begin{equation}
v^{2}\lambda ^{2}+(ab-(a+b)u +(u
^{2}+v^{2}))\lambda -\frac{1}{4}(a-b)^{2}=0.
\label{Eclambda1}
\end{equation}
whose solutions are conventionally denoted by  $\lambda _{1,2}(E,L)$ \ ($\lambda _{2}>\lambda _{1}$).
We notice that the first equation (\ref{dtdteta}) provides the dependence of time on the radial
coordinate r. The analytical result for this dependence is given in Appendix A.  By an inversion operation one obtains $r$ as a function of time.
Inserting this
in Eq.(3.5) and integrating the resulting equation one obtains $\theta$ as
function of time.
The two functions of time $r$ and $\theta$ are periodic. However, their
periods are different from each other. Moreover they are not commensurable which results in having
 open trajectories.
Here we give details about the description of the motion of the variable $r$,
the calculations for
the $\theta $ variable being performed by following similar steps.

For $r$, the period of the motion, $T _{U}(E,L)$, i.e. twice the time
elapsed between two successive passages through the turning points situated
at $\ r_{\min }=\sqrt{b(E,L)}$ and $\ r_{\max }=\sqrt{a(E,L)}$, is given by
the equation \cite{Pru,Whi,Rij}
\begin{equation}
T _{\text{U}}(E,L)=\pi \hbar \sqrt{\frac{3}{A^{\prime }F}}\frac{1}{%
\sqrt[4]{\Delta _{U}(E,L)}}\; _{2}F_{1}(\frac{1}{2},\frac{1}{2}%
;1;k_{U}^{2}(E,L)),
\label{TU}
\end{equation}

where
\begin{equation}
\Delta _{U}(E,L)=(ab-(a+b)u +(u^{2}+v%
^{2}))^{2}+(a-b)^{2}v^{2},
\label{DeltaU}
\end{equation}
with $a, b$ and $\rho$ depending on the energy E and the quantum number L  as shown in Eqs. (3.19) and (3.20).
$_{2}F_{1}(a,b;c;z)$ is the Gauss hypergeometric function and
\begin{equation}
k_{U}^{2}(E,L)=\frac{\lambda _{2}(E,L)}{\lambda _{2}(E,L)-\lambda _{1}(E,L)}.
\label{kU}
\end{equation}

The classical action corresponding to the degree of freedom $r$, written in units of $2\pi \hbar$
 is given by the following  integral:
\begin{equation}
I_{U}(E,L)=\frac{1}{2\pi \hbar }{\int}_{V_{eff \text{ min}2}(L)}^{E}T _{U}(w,L)dw.
\label{IU}
\end{equation}

\subsection{ Results for a "B" energy}
Similarly, in the case B, the zeros of the polynomial
$P_{4}(x;E,L)$, are positive numbers  denoted by
\begin{eqnarray}
x_{1}&=&a(E,L),\;x_{2}=b(E,L),\\ \nonumber
x_{3}&=&c(E,L),\;x_{4}=d(E,L),
\label{x1234}
\end{eqnarray}
and ordered as follows
\begin{equation}
x_{1}>x_{2}>x_{3}>x_{4}.
\end{equation}
The solution for $t$ as a function of $r$ is given in Appendix A.
The periods in both wells, left and right, are  equal to each other.
Indeed, one can prove that the two periods have a common expression:
\begin{equation}
T _{\text{B}}(E,L)=\pi \hbar \sqrt{\frac{3}{A^{\prime }F}}\frac{1}{%
\sqrt{(a(E,L)-c(E,L))(b(E,L)-d(E,L))}} {_{2}F_{1}}(\frac{1}{2},\frac{1}{2}%
;1;k_{B}^{2}(E,L)),
\label{TB}
\end{equation}
where
\begin{equation}
k_{B}^{2}(E,L)=\frac{(a(E,L)-b(E,L))(c(E,L)-d(E,L))}{%
(a(E,L)-c(E,L))(b(E,L)-d(E,L))}.
\label{kB}
\end{equation}
Since the hypergeometric function has a simple pole in $k=1$ the period of the motion diverges when the energy approaches
$V_{eff\text{ max}}(L)$ from below. If we calculate the period
for $E=V_{eff\text{ max}}(L)$, the same divergence is obtained.

For the left well, the integral action  given in units of $2\pi \hbar$ is equal to that
formulated for the right well but restricted to the corresponding energy interval:
\begin{equation}
I_{\text{left}}(E,L)=I_{B}(E,L)=\frac{1}{2\pi \hbar }
{\int}_{V_{eff\text{ \ min}1}(L)}^{E}T _{B}(w,L)dw.
\label{Ileft}
\end{equation}
The action for a trajectory of energy $E$ lying in the right well consists of
two terms:
\begin{equation}
I_{\text{right}}(E,L)=I_{U}(V_{eff\text{ min}
1}(L),L)+I_{B}(E,L).
\label{Iright}
\end{equation}
It turns out that $\ I_{U}(V_{eff\text{ min}1}(L),L)$ is an integer number, namely
\begin{equation}
I_{U}(V_{eff\text{ min}1}(L),L)=\frac{1}{2\pi \hbar }%
{\int}_{V_{eff\text{ min}2}(L)}^{V_{eff\text{ min}1}(L)}T_{\text{U}}(w,L)dw =L.
\label{IU1}
\end{equation}
Thus , the final expression for the integral action for a trajectory from the right well is
\begin{equation}
I_{\text{right}}(E,L)=L+I_{\text{left}}(E,L).
\label{Iright1}
\end{equation}
We notice that $k_{B}^{2}(E,L)$ $<$ $1$ for $E<V_{eff\text{ max}}(L)$, but $%
k_{B}^{2}(V_{eff\text{ max}}(L),L)$ $=$ $1$.

\subsection{Results for an "O" energy}

The case c) called "OVER" is similar to the case " UNDER". The only change
to be done in the formulae pertaining to case $U$ is the mere replacement $%
b(E,L)$ $\rightarrow $ $d(E,L)$ required by the conventional designations
of the zeros of the polynomial $P_4$.
Thus, the roots of $P_{4}(x;E,L),$ are
\begin{equation}
x_{1}=a(E,L),\;x_{2}=u(E,L)+iv(E,L)=x_3^*,\;
x_{4}=d(E,L),
\label{x1etc}
\end{equation}
with $x_{1}>x_{4}.$ (See also the notations for the  case b) BETWEEN, when all
roots are real and positive.)

Correspondingly, we denote by $\lambda _{1,2}(E,L)$ \ ($\lambda
_{2}>\lambda _{1}$) the solutions of the equation
\begin{equation}
v^{2}\lambda ^{2}+(ad-(a+d)u +u^{2}+v^{2})\lambda -\frac{1}{4}(a-d)^{2}=0.
\label{LamEq2}
\end{equation}
For a chosen energy in the interval O, the first equation (\ref{dtdteta}) has the 
analytical solution
given in Appendix A.

The period of the motion of the coordinate $r$, $\tau _{\text{O}}(E,L)$, i.e. twice the time
elapsed between two successive passages through the turning points situated
at $\ r_{\min }=\sqrt{d(E,L)}$ and $\ r_{\max }=\sqrt{a(E,L)}$, is given by
the equation

\begin{equation}
T _{\text{O}}(E,L)=\pi \hbar \sqrt{\frac{3}{A^{\prime }F}}\frac{1}
{\sqrt[4]{\Delta _{O}(E,L)}}\; _{2}F_{1}(\frac{1}{2},\frac{1}{2}%
;1;k_{O}^{2}(E,L)),
\label{TO}
\end{equation}

where
\begin{equation}
k_{O}^{2}(E,L)=\frac{\lambda _{2}(E,L)}{\lambda _{2}(E,L)-\lambda _{1}(E,L)}.
\label{kO}
\end{equation}
and $\Delta _{O}(E,L)$ is the discriminant of Eq. (\ref{LamEq2}).
When the energy approaches $V_{eff\text{ max}}(L)$ from above, $\lambda _{2}$ $%
\rightarrow $ $\infty ,$ and $k_{O}^{2}$ $\rightarrow $ $1.$ Thus, the
period of the motion diverges also when the energy approaches $V_{eff\text{ max}%
}(L)$ from above.

The integral action for the case O is:
\begin{equation}
\bigskip I_{\text{O}}(E,L)=I_{\text{right}}(V_{eff\text{ max}}(L),L)+%
\frac{1}{2\pi \hbar }{\int}_{V_{eff\text{ \ max}}(L)}^{E}T _{\text{O}}(w,L)dw.
\label{IO}
\end{equation}
\subsection{The virtual motion under the hump}
The region shown in Fig.1 under the hump is forbidden for the classical motion of r.
Indeed, there the energy is smaller than $V_{eff}$ which results in having a negative kinetic energy
as required by Eq.(\ref{rpunct}). Note that for this situation, the integral 
(\ref{dtdteta}) provides
imaginary values for time, i.e, $\tau=it$.
It is interesting to notice that by changing the time variable t by $-i\tau$
the resulting equation can be again integrated since the corresponding effective
potential is now $-V_{eff}$ while the energy becomes -E. Trajectories in the well
$-V_{eff}$ are conventionally called virtual. They are periodic curves in the
new variable $\tau$. Moreover this period may be connected with the tunneling time through the hump,
for the associated quantum system \cite{Hugo,Neg,Lan}.

The period of the motion in the well  $-V_{eff}$ $(L;r),$ or in the hump of
$V_{eff}$ $(L;r),$ is given by
\begin{equation}
T _{\text{virt}}(E,L)=\pi \hbar \sqrt{\frac{3}{A^{\prime }F}}\frac{1}{%
\sqrt{(a(E,L)-c(E,L))(b(E,L)-d(E,L))}} {_{2}F_{1}}(\frac{1}{2},\frac{1}{2}%
;1;k_{V}^{2}(E,L)),
\label{Tvirt}
\end{equation}

where
\begin{equation}
k_{V}^{2}(E,L)=\frac{(b(E,L)-c(E,L))(a(E,L)-d(E,L))}{%
(a(E,L)-c(E,L))(b(E,L)-d(E,L))}.
\label{kV}
\end{equation}

The integral action for the inverse of the hump potential is obtained in an analogous way with
the procedure described before. The final result is:
\begin{equation}
I_{\text{virt}}(E,L)=\frac{1}{2\pi \hbar }{\int}_{E}^{V_{eff\text{ }
\max }(L)}T _{\text{virt}}(w,L)dw.
\label{Ivirt}
\end{equation}
\subsection{Numerical application}
In order to emphasize on the complex structure of the classical motion
here we give
some relevant results for trajectories and their periods.
The coefficients used in our calculations are listed in Eq.(3.3). The shape of trajectories depend on the chosen value of energy. Here we consider a pair of trajectories corresponding to an energy lying close to  the maximum value of the effective potential, Fig. 2. One trajectory is lying in the left well, while the other one in the right well.
The two trajectories are quite different.
 Indeed, as shown in Fig. 2, in the left well $r$
 is running most of the distance $\sqrt{x_4}-\sqrt{x_3}$  in a very short time,
 than spends a large interval of time to reach the value $\sqrt{x_3}$ and to
 depart from it . Finally $r$ is coming quickly back to the value $\sqrt{x_4}$.
 The angle $\theta$ undertakes a jump of 2.5 $rad$  before ending a period of motion.
 As for the right well the $r$ curve corresponding to E close to $V_{eff\;max}$
is different from the one commented above. Indeed, Fig. 2 shows that $r$ stays
longer close to the value $\sqrt{x_2}$ and a very short time in the
neighborhood of $\sqrt{x_1}$. On the other hand, $\theta$ changes the speed around the half of
period which is at variance with the behavior of the trajectory from the left well.

From Fig. 2 we see that after an interval of two periods of $r$, 2T, the variable
$\theta$ covers about 11 $rad$ in the left well and only 5 $rad$ in the right well.
This suggests that the trajectory $r=r(\theta )$ is not closed. The two trajectories, 
lying in the left and right well respectively, correspond to E= 6.984326 MeV and are presented in Fig. 3.
As shown there, the inner and outer trajectories are almost tangent to each other. When the energy value is far from $V_{eff,max}$ there is an unreachable region, due to the bump.
The periods of trajectories
lying close to the maximum value of
$V_{eff}$ are much larger than those staying far from the top of the hump.
As a matter of fact, the period of $r$ has a discontinuity
for E=$V_{eff\;max}$, which is pictorially shown in Fig. 4. Such a discontinue behavior suggests that  $V_{eff\;max}$   is a critical point of a phase transition.
In one phase, for a given energy, the system may follow one of two $r$ trajectories of
equal periods, depending on the initial conditions,  while in the second phase only one trajectory is possible.

\section{The quantization of the periodic trajectories}
We note that the classical Hamiltonian given by Eq. (\ref{hasrond}) does not depend on the angle $\theta$
but only on its conjugate momentum. Consequently there is  a constant of motion
${\cal L}_3$ which is discretetized by the constraint (\ref{L3qua}) which in fact is a
way of quantizing it. The remaining variable r and its conjugate momentum $p_r$ are
both involved in the classical Hamiltonian. Moreover, as we have seen in the
previous Section, $r$ performs a periodical motion. Consequently, its motion
can be quantized by a constraint for the classical action similar to the Bohr-Sommerfeld
condition.
This condition will be applied separately for each region specified in Fig.1.
Thus, in the energy region called "UNDER", the quantization equation
\begin{equation}
I_{\text{U}}(E,L)=n_{1},
\label{IUeqn}
\end{equation}
gives the energy levels $E_{\text{under}}(L,n_{1})$ situated in the right
well below $V_{eff\text{ }\min 1}(L).$
It is remarkable that the above equation is satisfied by
\begin{equation}
E= V_{eff\text{ }\min 1}(L) \;\;, n_1=L.
\label{EeqV}
\end{equation}
This equation says that   $ V_{eff\text{ }\min 1}(L)$  is a quantal level
in both left and right wells. Moreover, below this energy, in the right well
 there are another L levels.
The quantization condition in the left and right  wells of the region "BETWEEN",
 reads
\begin{equation}
I_{\text{left}}(E,L)=n_{2} ,\;\;
I_{\text{right}}(E,L)=n_{1},
\label{Ileqn}
\end{equation}
These equations determine
 the same energy levels in both wells
\begin{equation}
E_{\text{left}}(L,n_{2})=E_{\text{right}}(L,n_{1}),
\label{EleqEr}
\end{equation}
where $n_{1}=n_{2}+L.$
The spectrum in the region "OVER" is obtained by solving the equation provided
by the quantization condition
\begin{equation}
I_{\text{O}}(E,L)=n_{3}.
\label{EOeqn}
\end{equation}
The virtual quantum states lying in the potential -$V_{eff}$ are obtained from
\begin{equation}
I_{\text{virt}}(E,L)=n_{4}.
\label{Ivirteqn}
\end{equation}
Having the integral action for the forbidden energy region, one can calculate the
transmission coefficient through the potential barrier.
Indeed, according to the WKB approximation, the transmission coefficient through the
hump is given by the equation:
\begin{equation}
D(E,L) \thickapprox  \exp (-2\pi I_{\text{virt}}(E,L)).
\label{DEL}
\end{equation}
\subsection{Numerical results}
Solving Eqs. (4.1), (4.3) and (4.5) satisfied by E, one finds the semiclassical spectrum of the system.
It is instructive to see how many quantum states with energy less than
$V_{eff\;max}$ accommodates the left well. The number of states is presented
in Fig. 5 as a function of angular momentum. Note that for some L, the number
of states is the same as for the L+1 case.  The quantal states are tentatively
organized in rotational bands in Fig. 6, in a full analogy with the standard
definition of the lowest bands within the liquid drop model. As for the right well the number of states and the rotational bands are shown in Figs. 7 and 8 respectively.
Figure 7 shows that for $L\leq 10$ the number of states of energy less then $V_{eff\;max}$ is 48, while for the next
12 angular momenta the number of states in the interval mentioned above  is 49.
Intervals for angular momenta for which the same number of states is registered is
decreasing and contains only one angular momentum  when the maximum for
the effective potential disappears. The "bands" in the right well differ the 
ones from the left well.
Indeed, as shown in Fig. 8, the energy spacing is much smaller then in the left well.
This is caused by the fact that the secondary minima of $V_{eff}$ are very close to each other, which is consistent with the known fact that  superdeformed nuclei exhibit larger moments of inertia.
Another difference between the two sets of bands consists of that while in the left well
the $n=2$  band is lower in energy than the $n=1$ band in the right well the ordering of the
two bands is opposite.

In contradistinction to what happens for real trajectories, the integral
action for the virtual states,
is a decreasing function of energy. Also, the number of states from $-V_{eff}$
is a decreasing function of angular momentum (see Fig. 9).

Energy levels for the virtual states do not coincide with energy levels corresponding to real states of left and right well.
The differences in energy and the period of virtual states satisfy the uncertainty relations.

Let us consider the transmission coefficients corresponding to the virtual
state of highest energy,
the first one under the hump. These are calculated within the WKB approximation by making use
of Eq.(47), for a given value of the angular momentum. The results are given
in Fig.10. It is worth to mention the discontinuity of D(E,L) given by Eq. (\ref{DEL})
for some values of L. Indeed, going from the values 10, 23, 29, 35, 40, 44, 47, 50, 53 to
11, 24, 30, 36, 41, 45, 48, 51, 54, respectively, one achieves a very big jump
in magnitude for the transmission coefficient. This result might have an important
consequence for the mechanism of alpha or heavy cluster emission. Indeed,
if the alpha's or heavy fragment would be in a state of angular momentum
staying in the list from above, its penetration probability would
be enlarged and a sizable emission rate could be registered.
Having in mind the integration limits in Eq.(4.7), the transmission
coefficient D(E,L), associated to the first virtual state depends on the difference
between  $V_{eff\;max}$ and the state energy. The smaller
the difference the larger the transmission coefficient.
Increasing the difference quoted above, the transmission coefficient decreases more or less
exponentially. Such a dependence may be seen in Fig. 11.
\section{Comparison with the previous work.}
Although the present and previous  work \cite{Rad2} use the same formalism to 
treat two distinct Hamiltonians, the results are both technically and physically different from each other. Thus, the fourth order boson  Hamiltonian used in Ref.\cite{Rad2} yields a classical effective potential which misses the superdeformed minimum.
Consequently, while here one distinguishes three distinct energy regions, called O, B and U, with topologically different trajectories, in the quoted reference there is only one region of finite trajectories and that correspond to the left well from Fig. 2. Here all trajectories  $r(t)$ are periodical. The period depends on energy. Indeed, increasing the energy from the U to the B region in the right well, there is a smooth variation for period. The periods associated to trajectories of the same energy but lying in the left and right well respectively, are equal to each  other. At the border between the B and the O regions the period
is singular, which suggests a phase transition. In the case of four order boson Hamiltonian such a transition appear to take place between a phase of periodic orbits and one of open orbits.
Although the periods for orbits of left  and right B trajectories of similar energies are equal to each other, the time dependence of $r$ and $\theta$ coordinates are different, respectively.  The difference is increasing when the energy approaches $V_{eff max}$.  As a matter of fact this reflects the influence of the deformation on trajectories. In this context one may say that in ref. \cite{Rad2} only moderately deformed trajectories show up which results in having in the plot corresponding to Fig. 3 only the inner curves. Thus, the superdeformed trajectories are specific to the present formalism. The absence of periodic orbits for $E>V_{eff max}$ has as consequence the fact that in a plot similar to that from Fig. 4, the descending branch  does not exist. 

Since for the chosen parameters the energy levels are almost equidistant one may expand the effective potentials around the minima and keep only the quadratic terms. The frequencies characterizing the harmonic motion in the two resulting wells have the expressions:
\begin{equation}
\omega_{k,L}=\left[3A'\left(\frac{4A'L^2}{r_{0,k}^4}+\frac{1}{3}A+Dr_{0,k}^2+\frac{5}{3}Fr_{0,k}^5\right)\right]^{1/2},\;\;k=1,2
\label{Omkl}
\end{equation}
where $r_{0,1}$ and $r_{0,2}$ denote the coordinates of the first and second minimum,  respectively. Note that for k=1, the influence of the sixth order boson term is the $r_{0,1}^5$ term in the above expression while the $k=2$ case exists only if the sixth order term is included. For the two wells one may define rotational bands in a similar way as in the framework of the liquid drop model.
\begin{eqnarray}
E^{(0)}_{k,L}&=&V^{min}_{eff}(L,r_{0,k}),\;\;L=0,2,4,...;k=1,2,
\nonumber\\
E^{(1)}_{k,L}&=&V^{min}_{eff}(L,r_{0,k})+\omega_{k,L},\;\;L=2,3,4,...;k=1,2,
\nonumber\\
E^{(2)}_{k,L}&=&V^{min}_{eff}(L,r_{0,k})+2\omega_{k,L},\;\;L=0,2,4,...;k=1,2.
\label{E012}
\end{eqnarray}

It is worth mentioning that the first equation (\ref{E012}) with k=1 represents a generalization of the variable moment inertia formula \cite{Mari,Hol}. When $V_{eff}$ is truncated to the quadratic term, the equation expressing the energy in terms of angular momentum derived in Ref.\cite{Hol} is obtained. The above cited equation considered for $k=2$ represents an extension of the variable moment of inertia formula to the superdeformed nuclei. The second and third equation
from (\ref{E012}) describe wobbling motions around the states from the ground band.
    
The results shown in Figs. 7, 8 are missing in ref. \cite{Rad2} since there, no right well in the effective potential shows up. Here we presented also the results characterizing the virtual states lying under the bump.
Also the transmission coefficients through the bump for a particular energy
has been studied as function of both angular momentum and the energy normalized to its maximum value.    

The peculiar feature of the model Hamiltonian studied so far consists of that it contains only a quadratic dependence on momenta which is consistent with the classical image where the kinetic energy involves only such type of terms. 
Moreover, six is the maximal order of boson terms which can be analytically treated. From this point of view the present treatment cannot be improved by adding higher order terms.

Formalisms dealing with anharmonic boson terms use Hamiltonians having, however, powers of momenta higher than two. Two examples of this type have been considered in Ref.\cite{Rad2}. Despite their complex structure, the chosen  boson Hamiltonians
were exactly solvable, producing finite formulae for energy as function of various quantum numbers. For the sake of saving the space here we consider only one of the two Hamiltonians mentioned above, corrected with a sixth order boson term:
\begin{equation}
H_2=\epsilon\sum_{\mu}b^{\dagger}_{\mu}b_{\mu}+\sum_{J=0,2,4}C_J\left[(   
b^{\dagger}b^{\dagger})_J(bb)_J)\right]_0+F(b^{\dagger}b^{\dagger})_0{\hat N}(bb)_0.
\label{H2}
\end{equation}
Here ${\hat N}$ denotes the quadrupole boson number operator (2.3). Using the same notations as in Ref.\cite{Rad2}, the average of $H_2$ on $|\Psi\rangle$ (2.5) has the expression:
\begin{eqnarray}
{\cal H}_2&=&\frac{A}{2}(q_1^2+q_2^2+p_1^2+p_2^2)+
\frac{B}{4}(q_1^2+q_2^2+p_1^2+p_2^2)^2+\frac{C}{8}(q_1p_2-q_2p_1)^2
\nonumber\\
&+&
\frac{F}{10}\left[\frac{1}{4}(q_1^2+q_2^2+p_1^2+p_2^2)^2-(q_1p_2-q_2p_1)^2\right](q_1^2+q_2^2+p_1^2+p_2^2).
\label{Hrond2}
\end{eqnarray}
As shown in Ref.\cite{Rad2}, ${\cal L}^2$ and ${\cal L}_3$ are constants of motion. Taking  the two constants equal to $\hbar^2L(L+1)$ and $\hbar M$ respectively, the energy  becomes a function of the quantum numbers $L$ and $M$:
\begin{eqnarray}
E_{LM}&=&2A\sqrt{L(L+1)}+4BL(L+1)+\frac{C}{2}M^2
\nonumber\\
&+&F\frac{8}{5}\sqrt{L(L+1)}\left[L(L+1)-M^2\right].
\label{ELM}
\end{eqnarray}
This energy is associated to the motion of the intrinsic degrees of freedom.
Supposing that these degrees of freedom are only weakly coupled to the Euler 
angles, then the total energy may be written as:
\begin{equation}
E_{JLM}=E_{LM}+\delta J(J+1)
\label{EJLM}
\end{equation} 
where J denotes the angular momentum in the laboratory frame.
On the other hand, the ground state energies can be obtained by averaging H on the states
$\{|Nv\alpha JM\rangle \}$ with $N=v=\frac{J}{2}$.
The same energies are obtained if in Eq. (\ref{ELM}) one substitutes:
\begin{equation}
2\sqrt{L(L+1)}\to\frac{J}{2},\;\; M=0
\end{equation}
Thus, energies of the ground band are given by the following equation:
\begin{eqnarray}
E_J&=&A\frac{J}{2}+B\frac{J^2}{4}+\delta J(J+1)+\frac{F}{5}\frac{J^3}{8}
\nonumber\\
   &\equiv& A_1J+B_1J^2+C_1J^3.
\label{JJ2J3}
\end{eqnarray}
The three parameters formula is expected to describe a larger number of experimental ground band energies than the two parameter formula from Ref. \cite{Rad2}.
To give an example we present, in Table 1, the results of a least square fit of the ground band energies for $^{232}$Th. From Table 1 one notices that the results provided by Eq.(\ref{JJ2J3}) are in much better agreement with the experimental data than those of Ref.[15].
\begin{table}[h]
\begin{tabular}{|c|c|c|c|c|c|c|c|}
\hline
$J^{\pi}$& Ex. & present&Ref.[15]&$J^{\pi}$&Ex. & present & Ref. [15]\\
\hline
$2^+$  & 0.049 &0.059&0.106  & $16^+$ &1.859  &1.849&1.823    \\
$4^+$  & 0.162 &0.175&0.246  & $18^+$ &2.263  &2.257&2.208    \\
$6^+$  & 0.333 &0.343&0.421  & $20^+$ &2.692  &2.693&2.629    \\
$8^+$  & 0.557 &0.561&0.632  & $22^+$ &3.144  &3.152&3.084    \\
$10^+$ & 0.827 &0.824&0.877  & $24^+$ &3.620  &3.631&3.574    \\
$12^+$ & 1.137 &1.129&1.158  & $26^+$ &4.116  &4.126&4.099    \\
$14^+$ & 1.483 &1.472&1.473  & $28^+$ &4.632  &4.633&4.659    \\
       &       &     &       & $30^+$ &5.162  &5.150&5.254    \\
\hline
\end{tabular}
\caption{Excitation energies obtained with Eq.(\ref{JJ2J3}) and with the two parameters formula derived in Ref.[15], respectively, and given in units of MeV, are compared with the corresponding experimental data \cite{Simon} for $^{232}$Th. The coefficients are determined by a least square procedure. The results for the coefficients involved in Eq(\ref{JJ2J3}) are:$A_1=14.96872$ keV 
$A_2=7.50384 keV$ and $A_3=-0.07602$ keV while the two parameters from the corresponding equation of Ref.[15] are:$A_1=44.01753$ keV, $A_2=4.37055$keV.}
\end{table}

Alternatively, the quantal energy for the intrinsic motion can be obtained by quantizing the anharmonic plane oscillator and the third component of angular momentum. In this case the result is:
\begin{equation}
E_{n,M}=A(n+1)+B(n+1)^2+\frac{C}{2}M^2+\frac{F}{5}\left[(n+1)^3-4(n+1)M^2\right].
\label{ENM}
\end{equation}
Neglecting the zero point energy terms and taking $M=\sqrt{L(L+1)}$ one obtains:
\begin{equation}
E_{n,L}=A_1n-B_1n^2+C_1L(L+1)+\frac{F}{5}\left(n^3-4nL(L+1)\right)
\label{ENL}
\end{equation}
with evident notations for the coefficients $A_1,B_1,C_1$.
From the above equation one sees that the sixth order term brings two types of corrections, one depending on $n^3$ and one depending on the product $nL(L+1).$
Equation (\ref{ENL}), involving four free parameters, seems to be very useful
to describe in a realistic fashion the excitation energies of states having the same angular momentum.  Extended calculations for ground band excitation energies with Eq.(\ref{JJ2J3}) as well as for excitation energies of a set of states of similar angular momentum with Eq.(\ref{ENL}) are in progress. The results will be published elsewhere.

\section{Summary and conclusions}
In the previous sections a sixth order boson Hamiltonian was semi-classically
studied.
The minimal action principle applied to the model Hamiltonian and a variational
state $|\Psi\rangle$ of a coherent type, provides a set of equations for the
complex parameters $z_0,z_2,z^*_0,z^*_2$ which play the role of classical
phase space coordinates.
In terms of polar coordinates ($r,\theta$), the classical equations of motion
have a canonical form.
There are two constants of motion and therefore the system is fully integrable.
The reduced classical phase space is one dimensional, the independent degree of freedom being
the radius $r$. For a suitable set of the Hamiltonian coefficients, the effective potential
involved in the classical energy function has a two minima shape.
The differential equation of $r$ can easily be integrated. The trajectories
$r(t)$ are periodic. Fixing the constant of motion ${\cal L}_3$ the
equation of motion
can be also analytically solved. The corresponding solution $\theta (t)$ is
also a periodical function of time.
The periods of $r(t)$ and $\theta (t)$ are different from each other. Moreover they are incommensurable and therefore
the trajectory $r(\theta)$ is not closed.
The emphasis is put mainly on the $r(t)$ trajectories. There are three distinct
energy regions, labeled by $U,B$ and $O$, where the $r(t)$ trajectories have distinct
behavior. Indeed, for a given energy E and a given value of L (half of ${\cal L}_3$)
 in the region U there is only one periodic trajectory, in B the are two trajectories of equal periods
 while in O again there exists only one periodic trajectory.
 The period magnitude depends on energy. It is an increasing function for
 $E<V_{eff\;max}$ and is singular at the interval end, while for the complementary region
 the period decreases when the energy is increased. This feature suggests that
$ E=V_{eff\;max}$ is a critical point where the system undertakes  a
phase transition.
Although the trajectories $r(t)$ have a complex structure which is far from that of a periodic harmonic
function, they can be quantized by a constraint for the integral action similar to the Bohr-Sommerfeld quantization condition.
In this way one obtains a semi-classical spectrum for each of the regions
$U,B$ and $O$. Choosing the constraint (\ref{L3qua}) for ${\cal L}_3$ in the region U there
are $L+1$ states while in B an equal number of states in the left and right wells, respectively.
The states in B are tentatively organized in rotational bands keeping close to
the conventional definition adopted by the liquid drop model.
The states of the right well are very much compressed compared with those of
the left well.
Moreover the $n=2$ band from the left well is lower in energy than the $n=1$
band while in the right band the bands ordering is opposite.

The tunneling phenomenon from the left to the right well through the separating barrier
was also studied. A special attention is paid to the first virtual state under the hump.
It is remarkable the fact that there  are several values for the angular momentum $L$ where
the transmission
coefficients, evaluated within the WKB approach, exhibit a spectacular
increase in magnitude. It is interesting to see whether this result has some
relevance for
alpha or heavy cluster emission. Indeed, if the preformed fragment  has an
 angular momentum equal to one of the
values where the transmission coefficients are very large one expects that the
emission rate is adequately increased. Represented as a function of the energy
distance between the chosen state and $V_{eff\;max}$ , the transmission
coefficient is
an exponentially decreasing function.

A detailed comparison of the present and Ref.\cite{Rad2}'s results is made
having the aim of pointing out the effect caused by the sixth order term.
In order to have a complete view of this comparison additional equations for 
the sixth order Hamiltonian have been derived: a)one giving the harmonic energies in the two wells of the effective potential b) a three parameters formula for the ground band excitation energies and c) an equation for energy describing the intrinsic motion, depending on a plane oscillator quantum number and an angular momentum associated to the boson degrees of freedom.

\section{Appendix A}
\renewcommand{\theequation}{A.\arabic{equation}}
\setcounter{equation}{0}
\label{sec:levelA}
Here we give the analytical solution of the first equation (\ref{dtdteta}) for an energy value
belonging to one of the three domains defined in Section III.
Aiming at a compact presentation of the final results it is useful to introduce the following notations:
\begin{eqnarray}
S_m&=&\frac{\hbar }{2}\sqrt{\frac{3}{A^{\prime }F}}\frac{1%
}{\sqrt[4]{\Delta _{m}}},\;\;m=U,O,\nonumber\\
S_{l}&=&\hbar \sqrt{\frac{3}{A^{\prime }F}}\frac{1}{%
\sqrt{(a-c)(b-d)}}=S_{r}.
\end{eqnarray}
Here $A',F$ are two of the Hamiltonian coefficients. $\Delta_m$- the second rank equation (\ref{Eclambda1},\ref{LamEq2}) discriminants and a,b,c,d are the roots of the fourth order polynomial $P_4(x;E,L)$ introduced in previous sections.
Two particular values of the variable $x$ are to be specified:
\begin{equation}
\mu _{m}=\frac{\lambda _{1}u+\frac{1}{2%
}(a+b)}{\lambda _{1}+1},\;\;
\eta _{m}=\frac{\lambda _{2}u+\frac{1}{2}%
(a+b)}{\lambda _{2}+1},\;\; m=U,O
\end{equation}
$\lambda_1$ and $\lambda_2$ are the solutions of Eqs. (\ref{Eclambda1}) and 
(\ref{LamEq2}) if m is equal to U and O, respectively. 
Also the following intervals for r are needed
\begin{eqnarray}
R^{(1)}_U&=&[\sqrt{b},\sqrt{\eta_U}],\;\;
R^{(2)}_U=[\sqrt{\eta_U},\sqrt{a}],\nonumber\\
R^{(1)}_O&=&[\sqrt{d},\sqrt{\eta_O}],\;\;
R^{(2)}_O=[\sqrt{\eta_O},\sqrt{a}],\nonumber\\
R^{(1)}_B&=&[\sqrt{d},\sqrt{c}],\;\;
R^{(2)}_B=[\sqrt{b},\sqrt{a}],
\end{eqnarray}
Integrating the first equation (\ref{dtdteta}) one obtains the time t as a function of r. In the intervals $R^{(1)}_m$ with $m=U,O$ and $R^{(i)}_B$ with i=1,2
the final result can be written in a compact form:
\begin{eqnarray}
t_m(E,L;r)&=&S_mF(\Phi_{1m}(E,L;r),k_n(E,L)),\;\;m=U,l,r,O,\nonumber\\
n&=&m(\delta_{m,U}+\delta_{m,O})+B(\delta_{m,l}+\delta_{m,r}).
\end{eqnarray}
If $r$ belongs to $R^{(1)}_m$ with $m=U,O$, the solution is:
\begin{equation}
t_{\text{m}}(E,L;r)=S_m\left( \frac{\pi }{2}\text{ }_{2}F_{1}(\frac{1}{%
2},\frac{1}{2};1;k_{m}^{2}(E,L))+F(\phi _{2\text{m}}(E,L;r),k_{m}(E,L))%
\right).
\end{equation}
$F(\Phi,k)$ denotes the incomplete elliptic integral of the first kind 
\begin{equation}
\bigskip F(\phi ,k)=\int_{0}^{\phi }\frac{d\theta }{\sqrt{%
1-k^{2}\sin ^{2}\theta }},
\end{equation}
The functions $\Phi_{1m}$ and $\Phi_{2m}$, $m=U,O$ have the expressions
\begin{eqnarray}
\phi _{1\text{m}}(E,L;r)&=& arccos\left( \frac{\eta _{m}-r^{2}}{%
r^{2}-\mu _{m}}\sqrt{\frac{\lambda _{1}(\lambda _{2}+1)}{%
\lambda _{2}(\lambda _{1}+1)}}\right) ,
\nonumber\\
\phi _{2\text{m}}(E,L;r)&=& arcsin\left( \frac{(\lambda
_{2}+1)(\lambda _{1}-\lambda _{2})}{\lambda _{2}(\lambda
_{1}+1)}\frac{1}{\left( \frac{r^{2}-\mu _{U}}{\eta _{U}-r^{2}}%
\right) ^{2}-\frac{\lambda _{2}+1}{\lambda _{1}+1}}\right) ^{\frac{%
1}{2}}, \;\; m=U,O.
\end{eqnarray}
For the $B$ energy region the functions $\Phi$ are
\begin{eqnarray}
\phi _{1l}(E,L;r)&=& arcsin\sqrt{\frac{%
(a-c)(r^{2}-d)}{(c-d)(a-r^{2})}}.
\nonumber\\
\phi_{1\text{r}}(E,L;r)&=& arcsin\sqrt{\frac{(a-c)(r^{2}-b)}
{(a-b)(r^{2}-c)}}.
\end{eqnarray}

%\end{document}
\newpage

\clearpage

\begin{figure}[h]
\centerline{\epsfig{figure=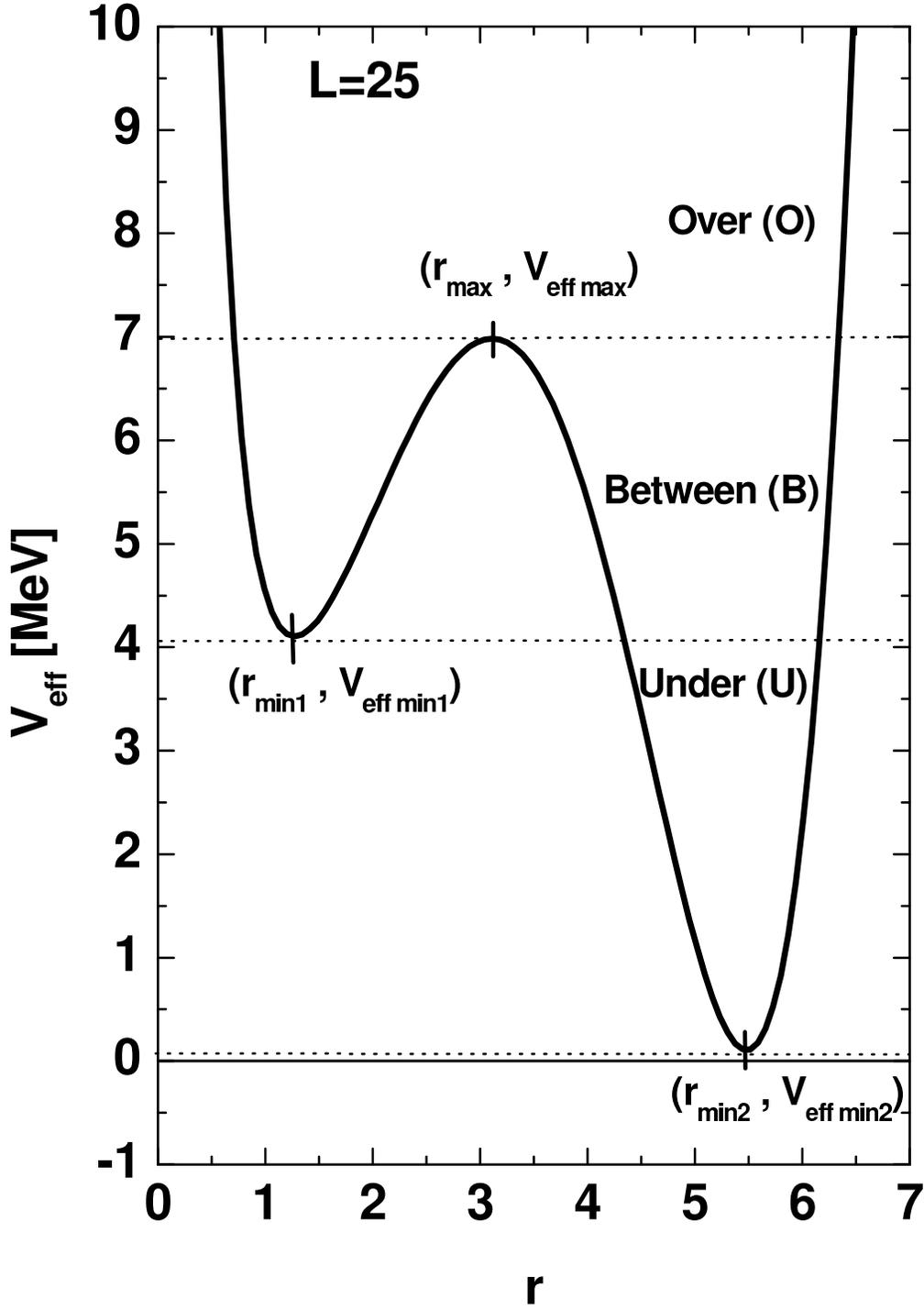,width=12cm,bbllx=3cm,%
bblly=10cm,bburx=18cm,bbury=26cm,angle=0}}
\vspace*{8cm}
\caption{The effective potential given by (\ref{Veff}) and (\ref{Vofr}) with the coefficients
$A',A,D,F$ specified in Eq. (\ref{coeff}), is plotted as function of the adimensional variable $r$ for  L=25 where L is defined by Eq. (L3).}
\label{Fig.1}
\end{figure}
\clearpage

\begin{figure}[h]
\begin{center}
\includegraphics[width=0.6\textwidth,angle=-90]{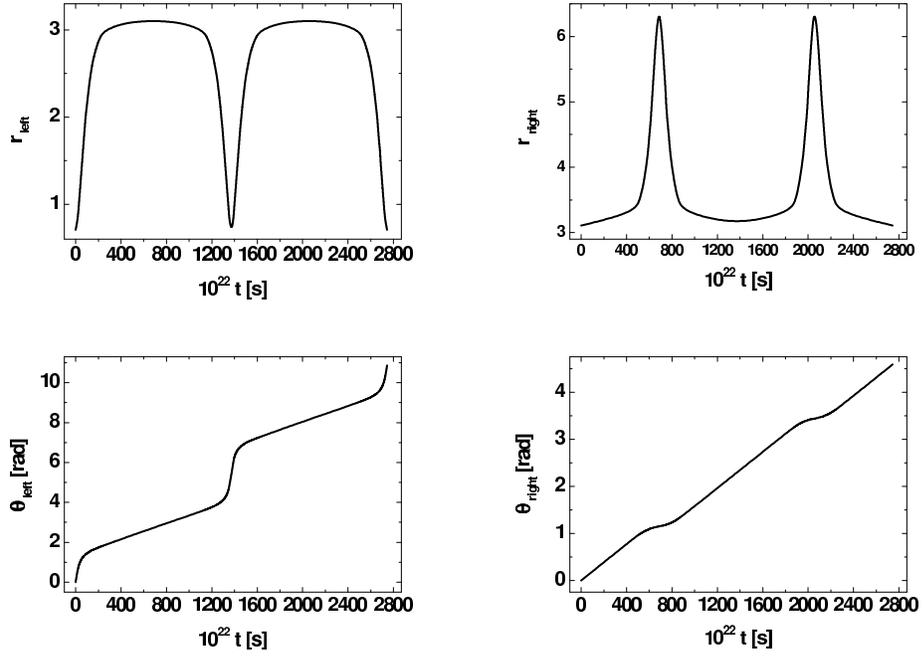}
\end{center}
%\centerline{\epsfig{figure=rrrfig4.eps,width=12cm,bbllx=5cm,%
%bblly=7cm,bburx=18cm,bbury=26cm,angle=-90}}
%\vspace*{10cm}
\caption{ The trajectories $r(t)$(adimensional) and $\theta(t)$(in units of $rad$) for the left and right wells,
E=6.984326 MeV and L=25, where L has the meaning specified by Eq. (\ref{L3qua}).}
\label{Fig.2}
\end{figure}

\clearpage

\begin{figure}[!ht]\
\begin{center}
\includegraphics[width=0.8\textwidth]{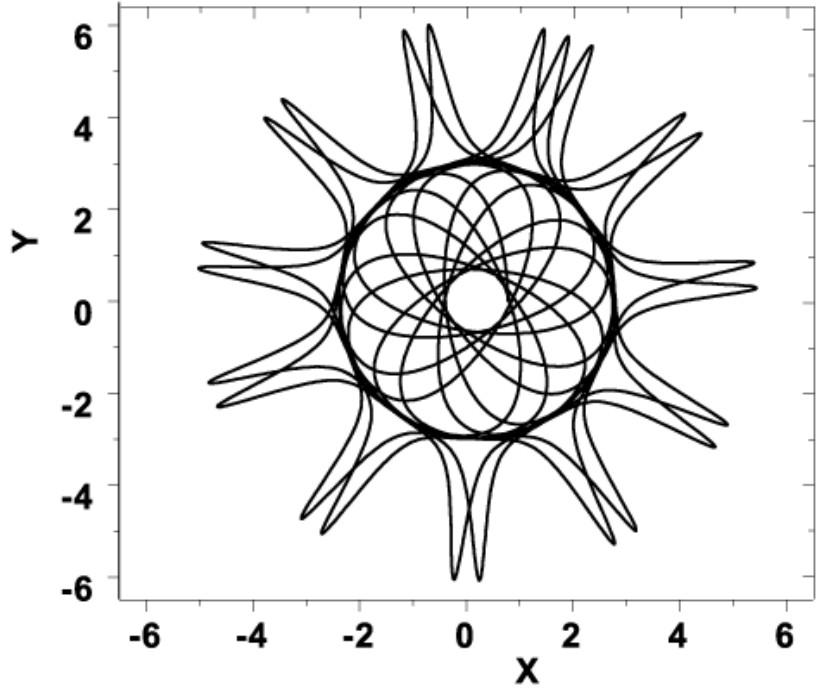}
\end{center}
\caption{The trajectories $r(\theta )$(adimensional) are represented in the Cartesian
coordinates
$x=r\rm{cos}(\theta)$ and $y=r\rm{sin}(\theta)$ for L=25 and E=6.984326 MeV. The
curves in both the left
(the inner curve)    and right
 (the outer curve) wells are given.}
\label{Fig.3}
\end{figure}

\clearpage

\begin{figure}[h]
\begin{center}
\includegraphics[width=0.6\textwidth,angle=-90]{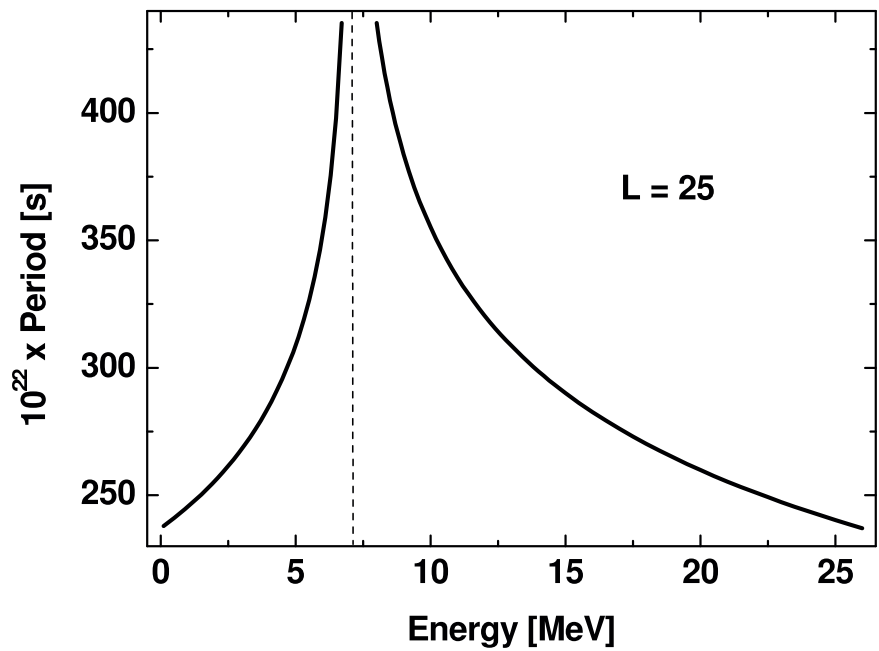}
\end{center}
%\centerline{\epsfig{figure=rrrfig7.eps,width=12cm,bbllx=5cm,%
%bblly=7cm,bburx=18cm,bbury=26cm,angle=-90}}
%\vspace*{10cm}
\caption{The trajectory periods in the left well of the region B as well as in
the region O
are plotted as function of energy.}
\label{Fig.4}
\end{figure}

\clearpage

\begin{figure}[h]
\begin{center}
\includegraphics[width=0.6\textwidth,angle=-90]{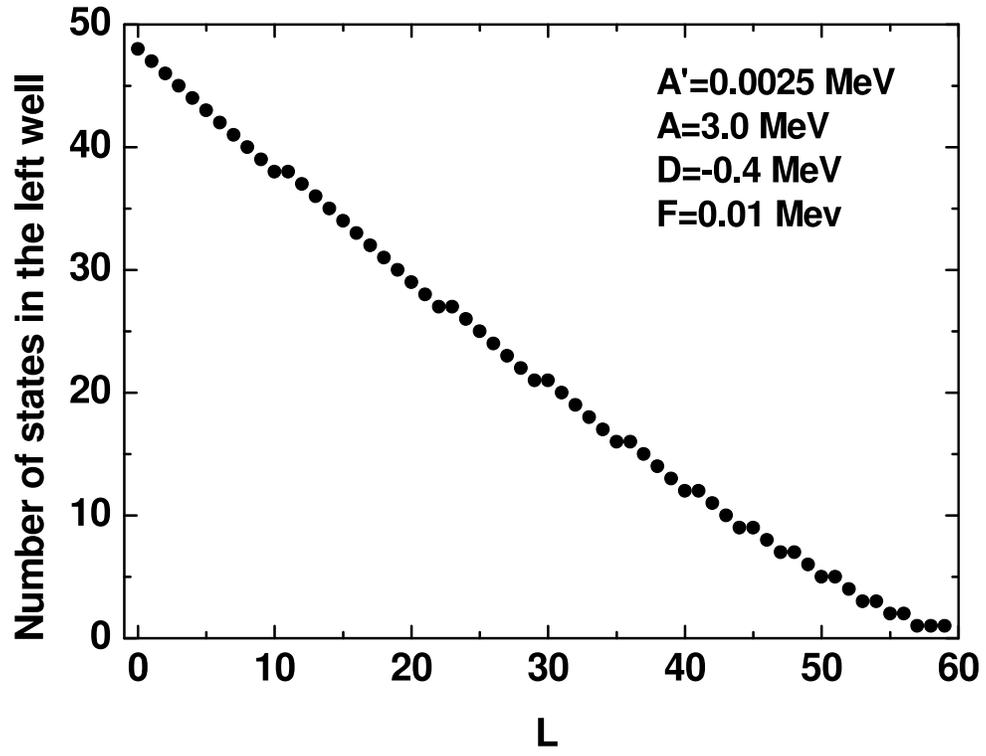}
\end{center}
%\centerline{\epsfig{figure=rrrfig9.eps,width=12cm,bbllx=3cm,%
%bblly=7cm,bburx=18cm,bbury=26cm,angle=-90}}
%\vspace*{9cm}
\caption{ The number of states from the left well are given as function of L 
(\ref{L3qua}).}
\label{Fig.5}
\end{figure}

\clearpage

\begin{figure}[h]
\begin{center}
\includegraphics[width=0.6\textwidth,angle=-90]{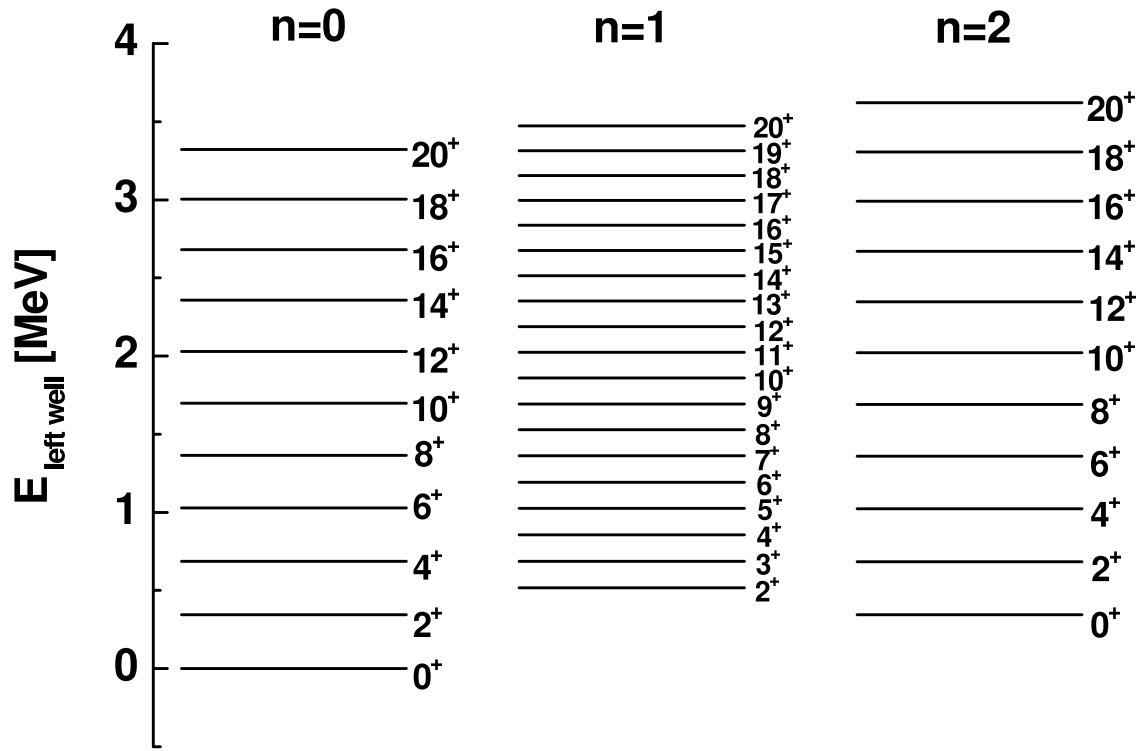}
\end{center}
%\centerline{\epsfig{figure=rrrfig10.eps,width=12cm,bbllx=3cm,%
%bblly=7cm,bburx=18cm,bbury=26cm,angle=-90}}
%\vspace*{9cm}
\caption{ The energy levels of the $n=0$, $n=1$ and $n=2$ bands from the left
well.}
\label{Fig.6}
\end{figure}

\clearpage

\begin{figure}[h]
\begin{center}
\includegraphics[width=0.6\textwidth,angle=-90]{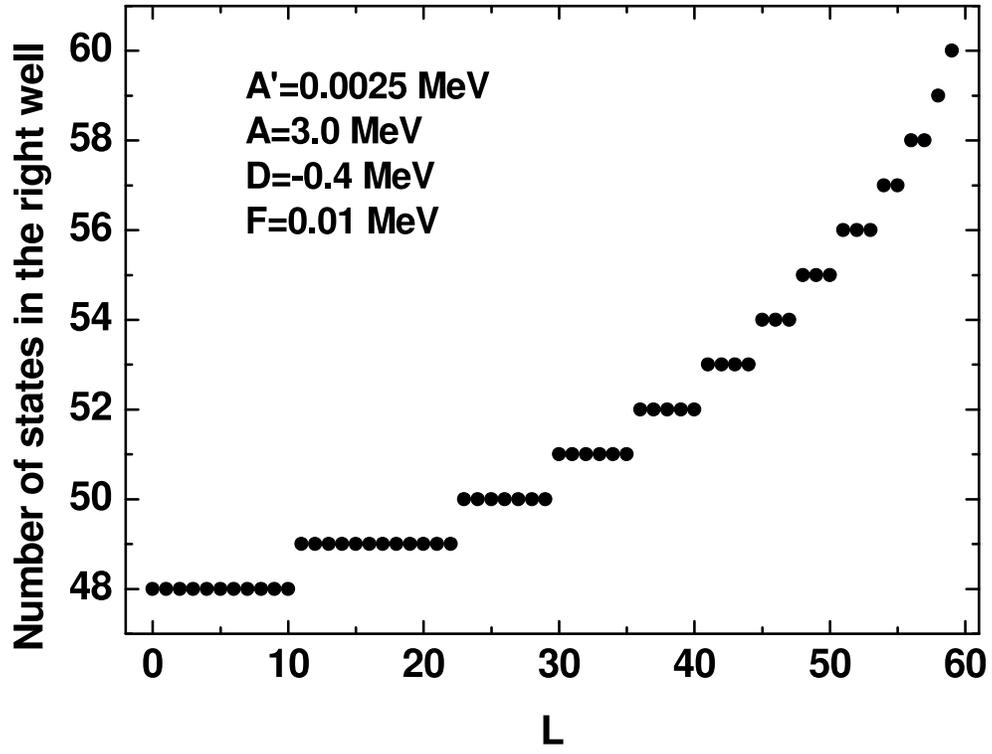}
\end{center}
%\centerline{\epsfig{figure=rrrfig12.eps,width=12cm,bbllx=3cm,%
%bblly=7cm,bburx=18cm,bbury=26cm,angle=-90}}
%\vspace*{9cm}
\caption{The number of states in the right well is plotted as function of 
L defined by Eq. (\ref{L3qua}).}
\label{Fig.7}
\end{figure}

\clearpage

\begin{figure}[h]
\begin{center}
\includegraphics[width=0.6\textwidth,angle=-90]{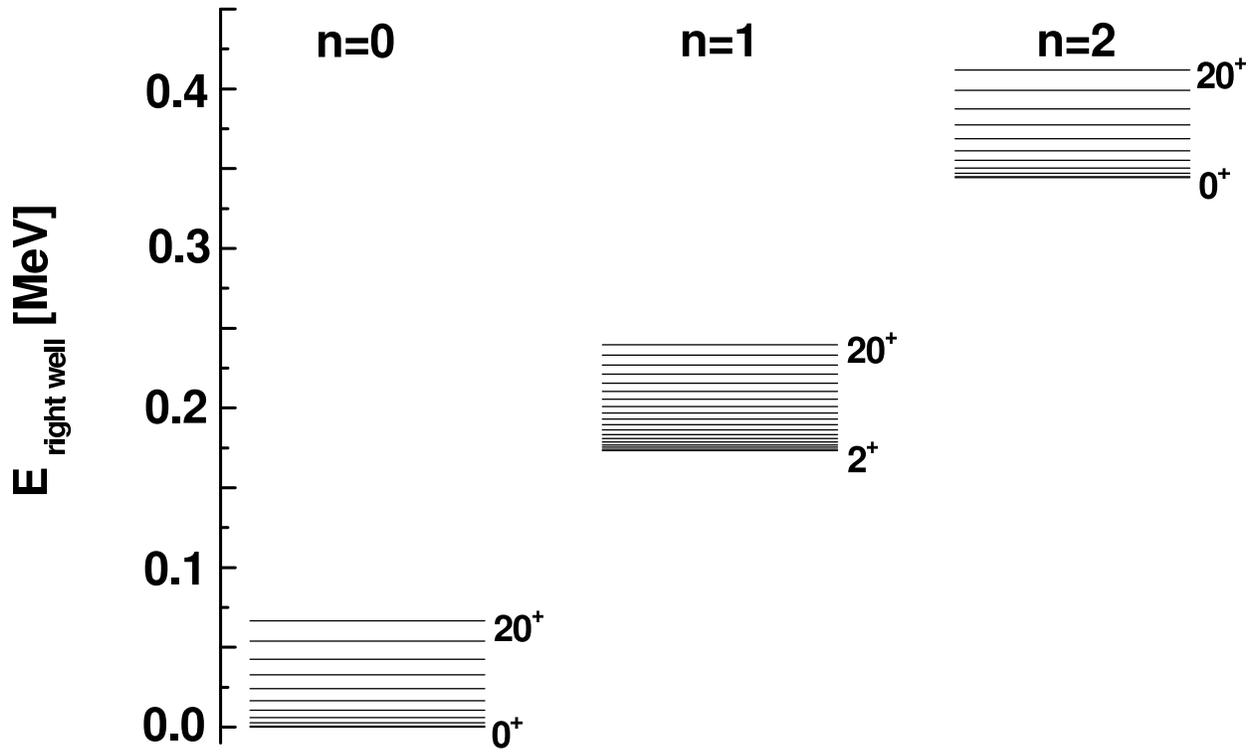}
\end{center}
%\centerline{\epsfig{figure=rrrfig13.eps,width=12cm,bbllx=3cm,%
%bblly=7cm,bburx=18cm,bbury=26cm,angle=-90}}
%\vspace*{9cm}
\caption{The energy levels in the $n=0$, $n=1$ and $n=2$ bands obtained in
the right well.
}
\label{Fig.8}
\end{figure}

\clearpage

\begin{figure}[h]
\begin{center}
\includegraphics[width=0.6\textwidth,angle=-90]{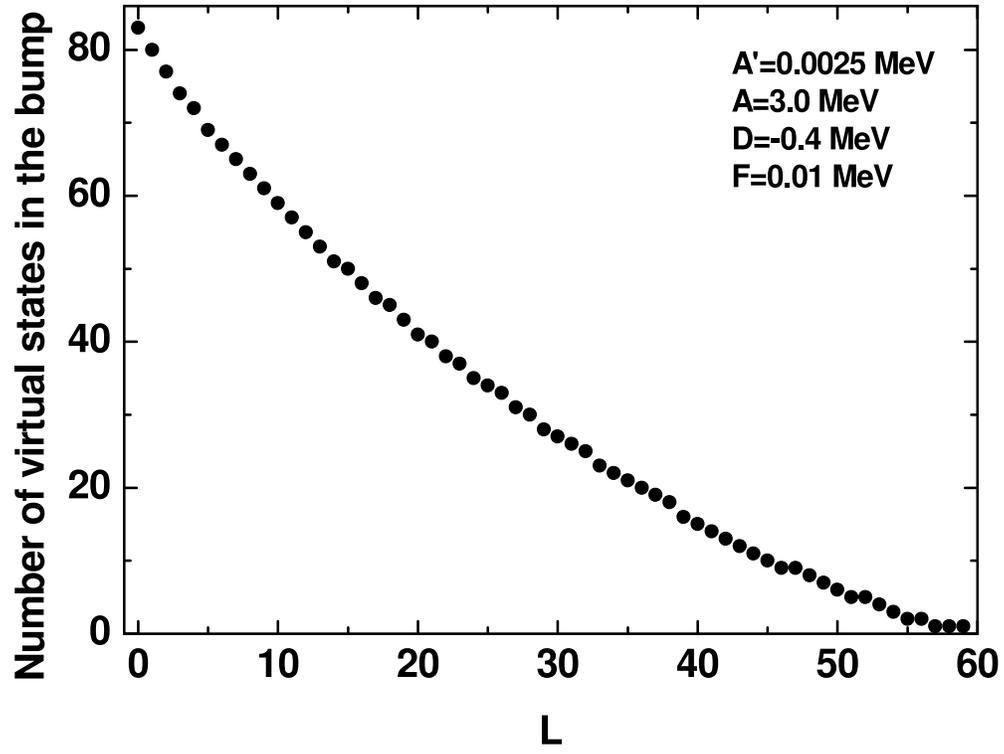}
\end{center}
%\centerline{\epsfig{figure=rrrfig15.eps,width=12cm,bbllx=3cm,%
%bblly=7cm,bburx=18cm,bbury=26cm,angle=-90}}
%\vspace*{9cm}
\caption{Number of virtual states which  appear under the effective potential
barrier.}
\label{Fig.9}
\end{figure}

\clearpage
\begin{figure}[h]
\begin{center}
\includegraphics[width=0.6\textwidth]{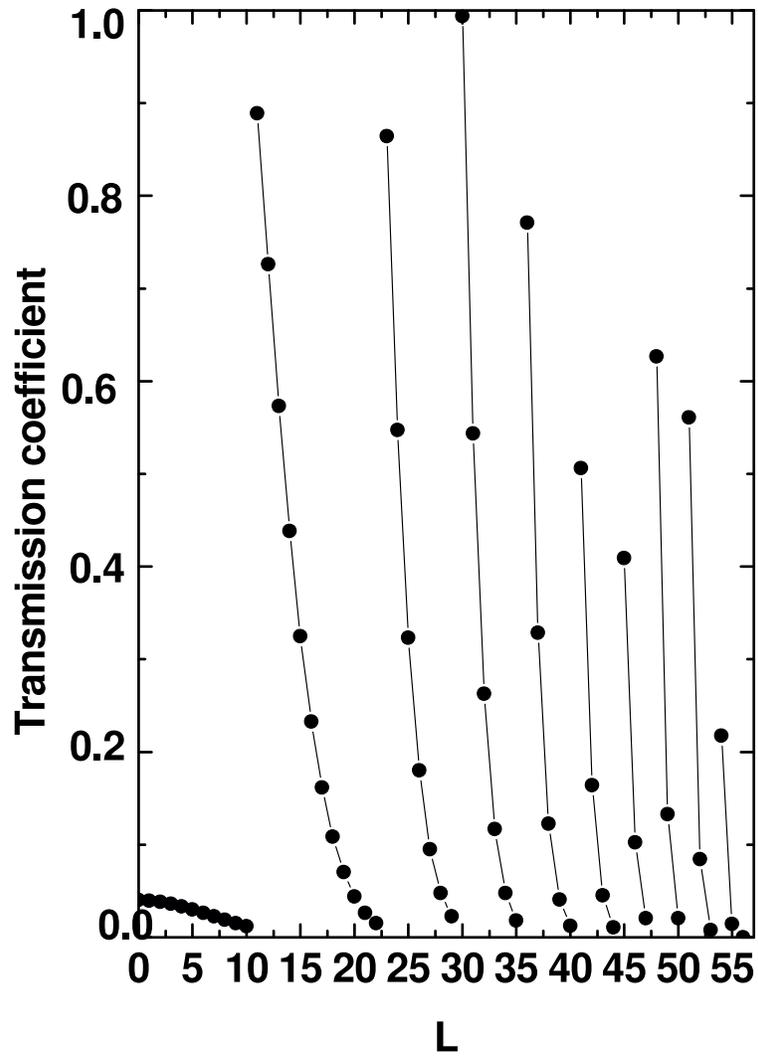}
\end{center}
%\centerline{\epsfig{figure=rrrfig16.eps,width=12cm,bbllx=3cm,%
%bblly=10cm,bburx=18cm,bbury=26cm,angle=0}}
%\vspace*{7cm}
\caption{The transmission coefficient calculated with Eq. (4.7) is plotted as
function of L given by Eq. (\ref{L3qua}).}
\vspace*{5cm}
\label{Fig.10}
\end{figure}

\clearpage
\begin{figure}[h]
\begin{center}
\includegraphics[width=0.6\textwidth,angle=-90]{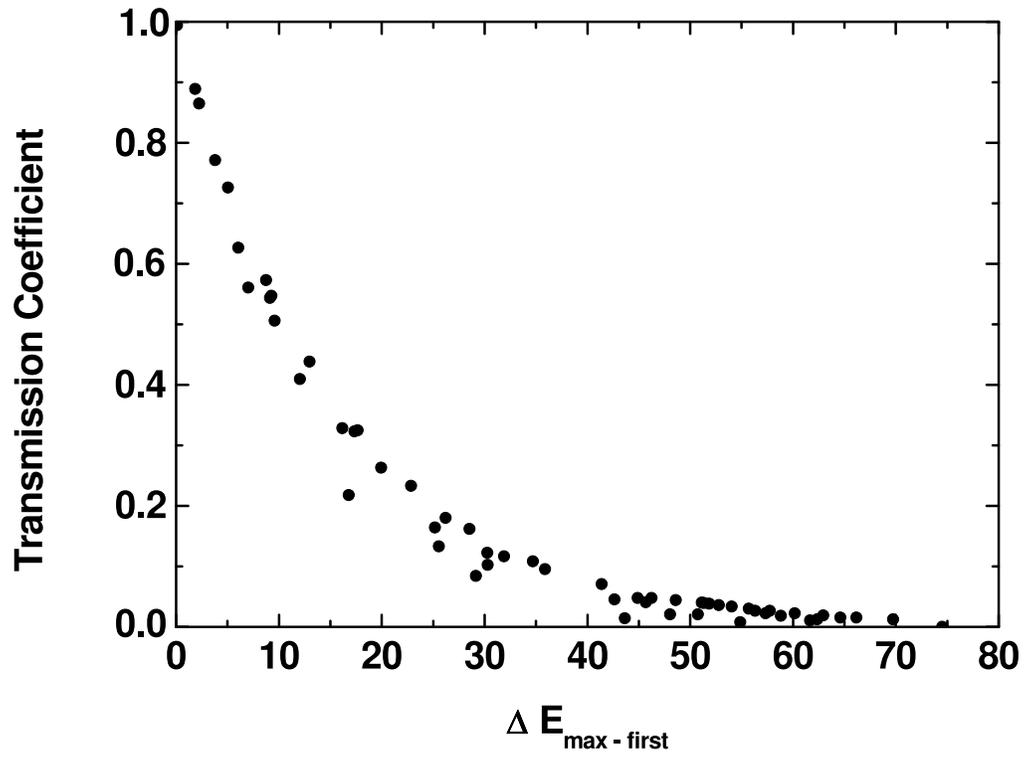}
\end{center}
%\centerline{\epsfig{figure=rrrfig18.eps,width=12cm,bbllx=3cm,%
%bblly=7cm,bburx=18cm,bbury=26cm,angle=-90}}
%\vspace*{9cm}
\caption{The transmission coefficient (adimensional) defined by Eq. (\ref{DEL}) is plotted vs.
$\Delta E_{first-max}$
 the excitation energy (given in units of $keV$) of the first virtual state.}
\label{Fig.11}
\end{figure}

\end{document}